\documentclass[twocolumn, pra, superscriptaddress,notitlepage]{revtex4}
\usepackage{graphicx}
\usepackage{physics}
\usepackage{epsfig}
\usepackage{color}
\usepackage{xspace}
\usepackage{amsmath}
\usepackage{amssymb}
\usepackage{amsthm}
\usepackage{amsfonts}
\usepackage{ulem}
\usepackage[dvipsnames]{xcolor}
\usepackage{bm}
\usepackage{bbm}
\usepackage[breaklinks,colorlinks,linkcolor=blue,citecolor=blue,urlcolor=blue]{hyperref}
\begin{document}
	
\global\long\def\id{\mathbbm{1}}
\global\long\def\ui{\mathbbm{i}}
\global\long\def\ud{\mathrm{d}}

\title{Measuring topological invariants of even-dimensional line-gapped non-Hermitian systems through quench dynamics}
\author{Xiao-Dong Lin}
\affiliation{Hefei National Research Center for Physical Sciences at the Microscale and School of Physical Sciences, University of Science and Technology of China, Hefei 230026, China}
\affiliation{School of Physics and Institute for Quantum Science and Engineering, Huazhong University of Science and Technology, Wuhan 430074, China}
\author{Long Zhang}
\email{lzhangphys@hust.edu.cn}
\affiliation{School of Physics and Institute for Quantum Science and Engineering, Huazhong University of Science and Technology, Wuhan 430074, China}
\affiliation{Hefei National Laboratory, Hefei 230088, China}

\begin{abstract} 
The accurate determination of non-Hermitian (NH) topological invariants plays a central role in the study of NH topological phases. In this work, we propose a general framework for directly measuring NH topological invariants in even-dimensional systems with real line gaps through quench dynamics. Our approach hinges on constructing an auxiliary Hermitian matrix topologically equivalent to the original NH Hamiltonian, enabling topological characterization via reduced-dimensional momentum subspaces 
called band-inversion surfaces (BISs).
A key insight lies in the emergence of chiral symmetry in the NH Hamiltonian specifically on BISs---a critical property that allows extension of the dynamical characterization scheme previously developed for odd-dimensional NH systems with chiral or sublattice symmetry [Lin {\it et al.}, Phys. Rev. Res. {\bf 7}, L012060 (2025)]. We show that NH topological invariants can be extracted from the winding patterns of a dynamical field constructed from post-quench spin textures on BISs. We demonstrate our approach through a detailed analysis of NH Chern insulators and then extend the framework to higher even-dimensional systems by introducing second-order BISs for characterization. 
{The framework is also generalized to imaginary line-gapped topological phases}.
This work establishes an experimentally accessible protocol for detecting NH topological invariants in quantum platforms.
\end{abstract}
\maketitle

\section{Introduction}

Non-Hermitian (NH) topological phenomena have attracted widespread attention in recent years due to their unique properties~\cite{Ashida2021_review,Bergholtz2021_review,Ding2022_review,Zhang2022_review,Lin2023_review,Okuma2023_review,Banerjee2023_review}, such as exceptional points (EPs)~\cite{Ding2022_review,Berry2004,Heiss2012} and the NH skin effect~\cite{Zhang2022_review,Lin2023_review,Okuma2023_review,Yao2018a,Borgnia2020,Okuma2020,Zhang2020}. 
Non-Hermiticity extends conventional topological classifications beyond the Hermitian framework and gives rise to novel topological phases~\cite{Gong2018,Kawabata2019a,ZhouH2019,LiuCH2019,Kawabata2019b,Yang2024,Leykam2017,Xu2017,Sun2021,Denner2021,Nakamura2024,Wojcik2020,LiZ2021,Hu2021,Liu2019,LeeCH2019,Luo2019,Manna2023}. To characterize NH topological phases, conventional topological invariants such as the winding number and Chern number have been generalized to NH systems~\cite{Lee2016,Lieu2018,Yin2018,Shen2018}, while new invariants such as vorticity have also been proposed~\cite{Leykam2017,Shen2018,Yin2018,Ghatak2019}. 
Under open boundary conditions, biorthogonal polarizations defined within the framework of biorthogonal quantum mechanics~\cite{Kunst2018,Edvardsson2019,Ghosh2022,Mandal2024} and non-Bloch topological invariants defined in the generalized Brillouin zone~\cite{Yao2018a,Yao2018b,Yokomizo2019,Yang2020,Wang2024,Xiong2024} have been introduced to restore the bulk-boundary correspondence. Therefore, the accurate determination of these NH topological invariants is central to
the investigation of NH topological phases.

Owing to their exceptional controllability,
quantum platforms, including ultracold atomic systems and solid-state spin systems, have emerged as ideal candidates for simulating diverse NH topological phenomena~\cite{Xu2017,Liu2019,Li2020,Zhou2021,Zhou2022}. Recent experimental advances include the realization of EPs and higher-order EPs~\cite{Li2019,Wu2019,Liu2021,Ren2022,WangC2024,Wu2024}, the exploration of nontrivial band braiding structures~\cite{Yu2022,WuY2023,Cao2023}, and the observation of the NH skin effect in one- and two-dimensional systems~\cite{Gou2020,Liang2022,Zhao2023}. 
These developments establish a critical foundation for investigating NH topological phases.
Nevertheless, direct experimental measurement of NH topological invariants remains a significant challenge, with only isolated successes reported~\cite{Su2021,ZhangW2021,Wang2021}. 

Recently, we proposed a general and unified framework for directly measuring various NH topological invariants through quench-induced non-unitary dynamics in odd-dimensional systems with chiral or sublattice symmetry~\cite{lin2025}.
The core methodology involves constructing an auxiliary Hermitian matrix 
$Q(\bm{k})$ that is topologically equivalent to the original NH Hamiltonian. This equivalence allows 
the topological characterization of NH systems to be mapped onto an equivalent problem for the Hermitian matrix $Q(\bm{k})$.
Our analysis demonstrates that, starting from a carefully selected initial state, the time evolution governed by a chiral/sublattice-symmetric NH Hamiltonian generates spin textures that encode the topological properties of $Q(\bm{k})$.
%Specifically, in systems with chiral symmetry, the time-averaged spin textures (TASTs) are found to be proportional to the components of the $Q(\bm{k})$ matrix. 
%such that $ \bm{g}(\bm{k}) = \hat{Q}(\bm{k})$, where $\hat{Q}(\bm{k})$ denotes the normalized $Q(\bm{k})$ matrix.
Accordingly, a dynamical field ${\bm g}({\bm k})$ can be constructed from these textures to characterize the NH topology. 
This approach provides an experimentally accessible route to determine NH topological invariants through measurable spin-texture dynamics.
However, the inherent symmetry constraints of this framework raise questions about its applicability to even-dimensional NH systems (e.g., NH Chern insulators~\cite{Shen2018,Kunst2018,Yao2018b,Chen2018,Hirsbrunner2019,Bartlett2023}), where such symmetries are typically absent.

In this work, we generalize the dynamical measurement scheme to even $d$-dimensional ($d$D)  {line-gapped} NH systems by leveraging the bulk-surface duality~\cite{Zhang2018,Zhang2019a,Zhang2019b,Yu2021}. 
This duality reduces the characterization of the equivalent  $d$D Hermitian $Q$ matrix to topological invariants defined on its 
$(d-1)$D momentum subspaces, known as band-inversion surfaces (BISs)~\cite{Zhang2018}. Crucially, for a broad class of NH systems with real line gaps, the NH Hamiltonian restricted to BISs exhibits emergent chiral symmetry, enabling the construction of a dynamical field from time-averaged spin textures (TASTs) to determine the NH topological invariants. 
We validate the scheme through numerical analysis of prototypical models in two and four dimensions.
This framework applies to even-dimensional NH topological phases with $\mathbb{Z}$ invariants classified within the NH Altland-Zirnbauer (AZ) or $\text{AZ}^{\dagger}$ symmetry classes~\cite{Kawabata2019a}.
We further demonstrate that the scheme can also be extended to systems with global sublattice symmetry and to imaginary line-gapped topological phases, as discussed in the Appendices.

The remainder of the paper is organized as follows. In Sec.~\ref{2DNH}, we investigate the dynamical measurement of topological invariants for two-dimensional (2D) NH Chern insulators. Beginning with a minimal model, we systematically illustrate the BIS-based characterization and subsequently extend the formalism to general cases.
In Sec.~\ref{HDNH}, we generalize this framework to higher even-dimensional systems, where second-order BISs are introduced to characterize NH topological invariants.
Conclusions and discussion are presented in Sec.~\ref{Conculsion}. Additional details are provided in the Appendices.

\section{Non-Hermitian Chern insulators}\label{2DNH}

We start with 2D NH Chern insulators, whose Hamiltonian generally takes the form:
\begin{align}~\label{NH Chern}
	\begin{split}
		H(\bm{k}) &= H_0(\bm{k})+\ui H_1(\bm{k}), \\
		H_0(\bm{k}) &= \sum_{i=x,y,z} h_i (\bm{k}) \sigma_i,\,\, H_1(\bm{k})=\sum_{i=x,y,z} h'_i (\bm{k}) \sigma_i
	\end{split}
\end{align}
where $\sigma_{x,y,z}$ are the Pauli matrices, $H_0(\bm{k})$ describes the Hermitian Chern insulator, and $H_1(\bm{k})$ introduces non-Hermiticity. Throughout main text, we restrict our discussion to real line-gapped regime.
The right eigenstates of $H(\bm{k})$ are denoted as $\ket{u^R_{n=\pm}(\bm{k})}$ with eigenvalues $E_\pm(\bm{k}) = \pm \varepsilon(\bm{k})$, 
where $\varepsilon(\bm{k})=[\sum_{i=x,y,z} (h_i+\ui h'_i)^2]^{1/2}$ for real line-gapped $H(\bm{k})$.

%we focus on real line-gapped cases, since only they exhibit nontrivial topology in the cases we are considering.

\subsection{A minimal model}\label{II.A}

We illustrate the scheme by first considering the minimal model with a single NH term.
Without loss of generality, we take $H_1(\bm{k}) = h'_z(\bm{k}) \sigma_z$, which serves as the basis for subsequent generalization.

To measure the NH topology of $H(\bm{k})$, we introduce an auxiliary Hermitian $Q$ matrix as
\begin{align}\label{Q_def}
	 Q(\bm{k})=\frac{1}{2}\left[\frac{H(\bm{k})}{\varepsilon(\bm{k})}+\frac{H^\dag(\bm{k})}{\varepsilon^*(\bm{k})}\right].
\end{align}
It can be demonstrated that $Q(\bm{k})$ is topologically equivalent to $H(\bm{k})$ in terms of line gaps~\cite{lin2025},
hereby mapping the topological characterization of NH systems to an equivalent Hermitian problem for $Q(\bm{k})$.
Crucially, based on the bulk-surface duality established in Refs.~\cite{Zhang2018,Zhang2019a,Zhang2019b,Yu2021}, this mapping allows further recasting the 2D Chern number that characterizes $Q(\bm{k})$ into a one-dimensional (1D) winding number defined on momentum subspaces called BISs. 

Specifically, we write $Q(\bm{k})=\sum_{i=x,y,z} h^Q_i(\bm{k})\sigma_i$, and define the BISs of $Q(\bm{k})$ along the $\sigma_z$-axis---the spin quantization axis aligned with the NH term. The BISs then correspond to momenta where $h^Q_z(\bm{k})=0$, and the topology of $Q(\bm{k})$ is characterized by the winding of the normalized spin-orbit coupling field ${\bm \hat{\bm{h}}_{\rm so}^{Q}}(\bm{k})=(\hat{h}^Q_x(\bm{k}), \hat{h}^Q_y(\bm{k}))$ along all BISs~\cite{Zhang2018}:
\begin{equation}~\label{WQ}
	W_Q=\sum_j\frac{1}{2\pi}\int_{{\rm BIS}_j} d{\bm{k}}\left(\hat{h}^Q_x\partial_{\bm{k}} \hat{h}^Q_y - \hat{h}^Q_y\partial_{\bm{k}}\hat{h}^Q_x\right),
\end{equation}
where $\hat{h}^Q_i =h^Q_i/\sqrt{ (h^Q_x)^2+(h^Q_y)^2}$ ($i=x,y$) and ``${\rm BIS}_j$'' denotes the $j$th BIS.

A key advantage of this BIS-based characterization lies in the emergent chiral symmetry of the dimensionally reduced NH Hamiltonian $H(\bm{k})$ defined on the BISs of $Q(\bm{k})$: $\sigma_z H^\dag({\bm k})\sigma_z^{-1}= -H({\bm k})$.
This symmetry arises because the condition defining the BISs, $h_z^Q(\bm{k}) = 0$, is equivalent to $h_z(\bm{k}) = 0$ (see Appendix~\ref{LocateBISs}).
The emergent chiral symmetry underpins the dynamical measurement protocol in Ref.~\cite{lin2025}, 
enabling the direct extraction of the topological invariant $W_Q$ from quench dynamics.

%This definition of BISs ensures chiral symmetry of the NH Hamiltonian $H(\bm{k})$ on the BISs, as they coincide with the momentum subspace satisfying $h_z(\bm{k}) = 0$, and the $h_i^Q(\bm{k})$ is clearly expressed as $h_i^Q(\bm{k}) =   h_i (\bm{k})/\varepsilon(\bm{k})$ with $\varepsilon(\bm{k}) = \sqrt{\sum_{i=x,y} h_i^2(\bm{k})-{h'}^2_z(\bm{k})}$ on the BISs. The chiral symmetry is essential for the dynamical classification approach proposed in~\cite{lin2025}, which enables the extraction of the topological invariant $W_Q$ from quench dynamics. 

Following Ref.~\cite{lin2025}, we investigate the TASTs after a quench:
\begin{align}\label{TAST}
	\langle\sigma_i(\bm{k})\rangle_T=\lim_{T\to \infty}\int_0^T\frac{1}{{\cal N}_{\bm{k}}}{\rm Tr}\left[\rho_0(\bm{k},t)\sigma_i\right] dt,\,\, i=x,y,z
\end{align} 
where $\rho_0(\bm{k},t) = e^{-\ui H(\bm{k})t}\rho_0(\bm{k}) e^{\ui H^{\dagger}(\bm{k})t}$ evolves from the initial state $\rho_0(\bm{k}) = \ket{\psi_0(\bm{k})}\bra{\psi_0(\bm{k})} $ with ${\cal N}_{\bm{k}}\equiv{\rm Tr}\left[\rho_0(\bm{k},t)\right]$ being the normalization factor. 
Note that these TASTs are defined solely in the basis of right eigenvectors, enabling
straightforward experimental implementation.
One can prove that the BISs can be dynamically identified by 
\begin{equation}\label{dBIS}
\langle\sigma_z(\bm{k})\rangle_T = 0,
\end{equation}
which holds for an arbitrary initial state (see Appendix~\ref{LocateBISs} for proof).
A dynamical field $\bm{g}(\bm{k})=(g_x(\bm{k}),g_y(\bm{k}))$ can be defined from the normalized TASTs through
\begin{equation}\label{g_def}
	\begin{split}
		g_{x}(\bm{k}) = \overline{\langle\sigma_y(\bm{k})\rangle}_T,\quad
		g_{y}(\bm{k}) =-\overline{\langle\sigma_x(\bm{k})\rangle}_T,
	\end{split}
\end{equation}
where $\overline{\langle\sigma_i(\bm{k})\rangle}_T={{\langle\sigma_i(\bm{k})\rangle}_T}/{\sqrt{\sum_{j=x,y} \langle\sigma_{j}(\bm{k})\rangle_T^2}}$ ($i=x,y$).
By choosing the initial state to be polarized along the chiral symmetry axis (aligned with the NH perturbation), i.e., $\rho_0=(\id+\sigma_z)/2$~\cite{note1}, the constructed dynamical field satisfies $g_i(\bm{k})\vert_{{\bm k}\in \mathrm{BIS}}\approx\hat{h}^Q_i(\bm{k})$~\cite{note2} (see Appendix~\ref{measure}). 
Thus, the topology of $Q(\bm{k})$ can be dynamically measured through the winding of $\bm{g}(\bm{k})$ on BISs:
\begin{equation}\label{W1}
	W_1=\sum_j\frac{1}{2\pi}\int_{{\rm BIS}_j} d{\bm{k}}\left(g_x\partial_{\bm{k}} g_y - g_y\partial_{\bm{k}}g_x\right).
\end{equation}
Equations~\eqref{dBIS} and \eqref{W1} provide a direct measurement of the real line-gapped topology of $H(\bm{k})$.

\begin{figure}
	\includegraphics[width=0.48\textwidth]{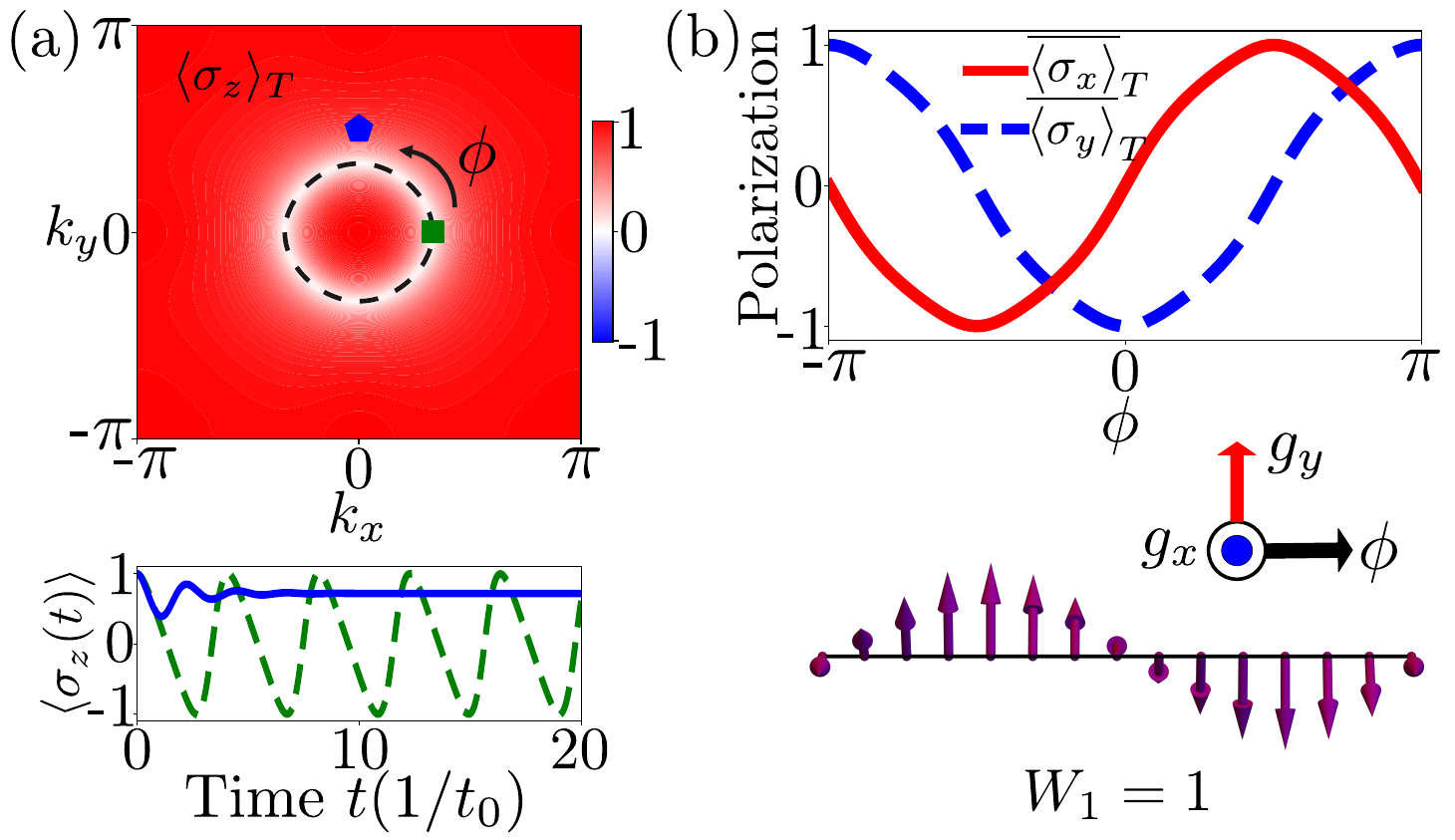}
	\caption{Measuring the topology of the minimal NH QAH model. (a) A ring-shaped BIS (black dashed curve) is identified via $\langle \sigma_z(\bm{k}) \rangle_T = 0$. The lower panel shows how the spin polarization $\langle \sigma_z(t) \rangle$ evolves at a momentum on (green) or off (blue) the BIS. (b) 
Upper panel: The normalized TASTs $\overline{\langle \sigma_{x,y}\rangle}_T$ along the BIS as functions of the azimuthal angle $\phi$.
Lower panel: The constructed dynamical field $\bm{g}(\phi)$ (purple arrows) exhibits a nontrivial winding $W_1 = 1$. Here we take the parameters $(m_z,t_{so},\eta)= (3,0.5,0.4)t_0$ and the initial state $\rho_0 = (\id+\sigma_z)/2$.}\label{fig1}
\end{figure}

As a paradigmatic example, we investigate a NH extension of the quantum anomalous Hall (QAH) model, described by the Hamiltonian
\begin{equation}
	\begin{aligned}
		H_0(\bm{k}) =&\,2t_{\rm so}\sin k_x \sigma_x + 2t_{\rm so} \sin k_y \sigma_y\\
		&+ (m_z-2t_0 \cos k_x-2t_0\cos k_y)\sigma_z,\\
		H_1= &\, \eta \sigma_z,
	\end{aligned}
\end{equation}
where $t_0$ ($t_{\rm so}$) represents the spin-conserved (spin-flipped) hopping coefficient, and $m_z$ ($\eta$) denotes the real (imaginary) Zeeman field. The Hermitian QAH model $H_0$, having been experimentally realized in optical Raman lattices~\cite{Zhan2016,Sun2018a,Liang2023}, exhibits topologically nontrivial phases within the parameter regime $0<|m_z|<4t_0$, characterized by a Chern number $C={\rm{sgn}}(m_z)$. The NH perturbation $H_1$ can be achieved by a spin-dependent atom loss induced via a resonant coupling to short-lived excited states~\cite{Ren2022}.
Numerical results are presented in Fig.~\ref{fig1} with parameters $(m_z, t_{\rm so}, \eta)= (3,0.5,0.4)t_0$.
As shown in Fig.~\ref{fig1}(a), a ring-shaped BIS (black dashed curve) is identified via the vanishing time-averaged spin polarization $\langle \sigma_z (\bm{k}) \rangle_T=0$.
Notably, the NH Hamiltonian exhibits strictly real spectra on the BIS, where quench dynamics preserve unitarity---a characteristic analogous to Hermitian systems.
Consequently, the spin polarization oscillates persistently around zero on the BIS, resulting in a vanishing time-averaged polarization.
In contrast, regions away from the BIS feature complex energy spectra, causing rapid decay of spin oscillations toward non-vanishing expectation values $\langle u_\pm^R(\bm{k}) | \sigma_z | u_\pm^R(\bm{k}) \rangle$ [Fig.~\ref{fig1}(a), lower panel].
Figure~\ref{fig1}(b) displays the normalized TASTs and corresponding dynamical field ${\bm g}(\phi)$ (purple arrows) along the BIS, parameterized by the azimuthal angle 
$\phi\in(-\pi,\pi]$. The nontrivial winding of ${\bm g}(\phi)$ yields a quantized topological invariant $W_1 = 1$, demonstrating the persistence of $H_0$'s topology under NH perturbations.

%The reason that one can determine the BISs from $\langle\sigma_z(\bm{k})\rangle_T = 0$ stems from the fundamental connection between TASTs and the system's eigenvalue $E_n(\bm{k})$. For real spectrum ${\rm Im}(E_n(\bm{k})) = 0$, the quench dynamics remain unitary, rendering the situation analogous to the Hermitian systems, while for complex spectrum, after long-term dynamics, only the right eigenstates with ${\rm Im}(E_n(\bm{k})) > 0$ persist, TASTs converge towards the spin texture of eigenstates. On the BISs, the NH Hamiltonian exhibits real spectrum, the spin textures exhibits persistent oscillations around zero, leading to a vanishing time average, as indicated by the green dashed curve in the lower panel of Fig~\ref{fig1}(a). Away from the BISs, the NH Hamiltonian has complex spectrum, the spin textures decay and the TASTs approach $\langle u_n^R(\bm{k}) | \sigma_z | u_n^R(\bm{k}) \rangle$, as shown by the blue curve. 
%
%The topological characterization is completed by analyzing the dynamical field $\bm{g}(\bm{k})$, which is derived from the normalized TASTs $\overline{\langle\sigma_{x,y}(\bm{k})\rangle}_T$. Fig~\ref{fig1}(b) presents the normalized TASTs and the corresponding dynamical field along the BISs, which is parameterized by the angle $\phi$. The nontrivial winding pattern of $\bm{g}(\bm{k})$ shown as purple arrows yields a quantized winding number $W_1 = 1$, confirming the survival of topology under non-Hermiticity.

\subsection{General scenarios}\label{II.B}

\begin{figure}
	\includegraphics[width=0.48\textwidth]{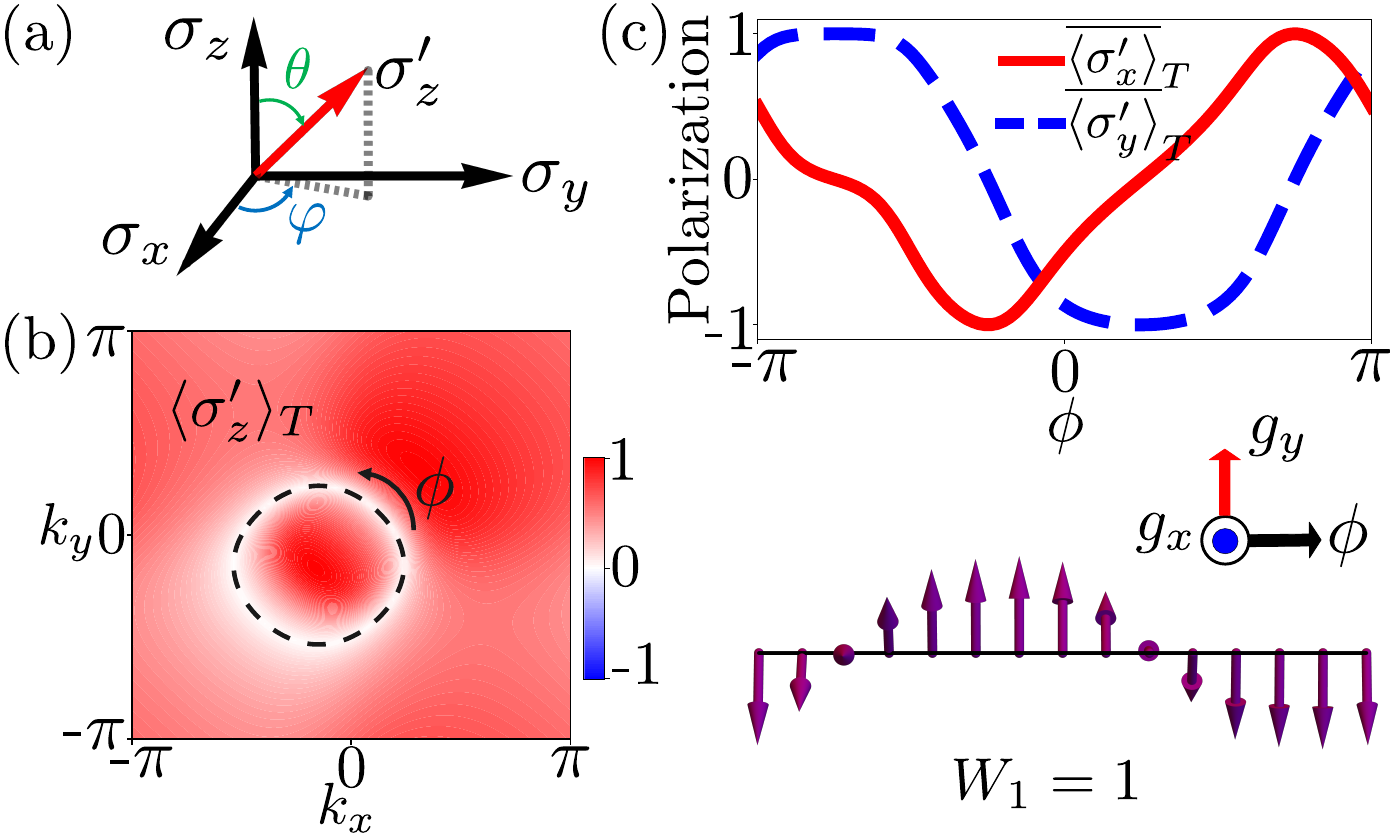}
	\caption{ Measuring the topology of the generalized NH QAH model. 
	 (a) A spin rotation is performed to align the transformed $\sigma'_z$-axis with the direction $(\theta, \varphi)$ determined by the NH perturbations.
	 (b) A ring-shaped BIS (black dashed curve) is identified via the rotated TAST $\langle \sigma'_z(\bm{k}) \rangle_T = 0$. 
	 (c) Upper panel: The normalized TASTs $\overline{\langle \sigma'_{x,y}\rangle}_T$ along the BIS as functions of the azimuthal angle $\phi$.
 	 Lower panel: The constructed dynamical field $\bm{g}(\phi)$ (purple arrows) exhibits a nontrivial winding $W_1 = 1$.
	 Here we take the parameters $(m_z,t_{so},\eta)= (3,0.5,0.3)t_0$ and the initial state $\rho_0 = (\id+\sigma'_z)/2$.}\label{fig2}
\end{figure}

The above scheme can be extended to more general scenarios, where the NH perturbation $H_1(\bm{k})=\sum_i h'_i(\bm{k})\sigma_i$ incorporates multiple NH terms while maintaining a fixed spinor orientation $(\theta, \varphi)$ in the space spanned by the Pauli matrices $\sigma_{x,y,z}$, as depicted in Fig. \ref{fig2}(a). 
In such cases, a spin rotation can be implemented to align the transformed $\sigma'_z$-axis with the predetermined direction $(\theta, \varphi)$. This rotation is formally expressed as
\begin{align}\label{Rotation}
	\sigma_i' = \sum_{j=x,y,z}R_{ij} \sigma_j, \quad i=x,y,z,
\end{align}
where $R_{ij}$ denotes the $(i,j)$th element of the $3\times3$ rotation matrix $R(\theta,\varphi)$, with $\theta$ and $\varphi$ parametrizing the rotation angles. In the rotated frame, the NH Hamiltonian becomes $H(\bm{k}) = \sum_{i=x,y,z} h_i^r(\bm{k}) \sigma'_i + \ui {h'}_z^r(\bm{k})\sigma'_z $, which takes the same form as the minimal model. The transformed vector field $\bm{h}^r(\bm{k})\equiv(h_x^r(\bm{k}),h_y^r(\bm{k}),h_z^r(\bm{k}))$ relates to the original $\bm{h}(\bm{k})\equiv(h_x(\bm{k}),h_y(\bm{k}),h_z(\bm{k}))$ through the rotation 
$h_i^r(\bm{k}) = \sum_j h_j(\bm{k})R_{ji}$.

Dynamical characterization proposed in Eqs.~\eqref{dBIS} and \eqref{W1} can be accordingly reestablished in the rotated frame. The BISs, defined as momentum subspaces where $h_z^r(\bm{k}) = 0$, are determined by the condition $\langle \sigma'_z(\bm{k}) \rangle_T=0$. The topology of $H(\bm{k})$ is then characterized by the winding number of ${\bm g}(\bm{k})$ along the BISs, which is constructed from the rotated TASTs $\langle \sigma'_{x,y}(\bm{k}) \rangle_T$ with the initial state $\rho_0=(\id+\sigma'_z)/2$. Notably, $\langle \sigma_i'(\bm{k})\rangle_T$ can be derived directly from $\langle \sigma_i(\bm{k})\rangle_T$ through the transformation
\begin{equation}\label{rotateTASTs}
	\langle \sigma_i'(\bm{k}) \rangle_T =\sum_{j=x,y,z}R_{ij}\langle \sigma_j(\bm{k}) \rangle_T,\quad i=x,y,z,
\end{equation}
enabling the extraction of the rotated observables via straightforward data post-processing.
As spin textures in the original frame are directly accessible via quantum quench protocols, 
the dynamical characterization scheme is experimentally implementable for all scenarios discussed here.

A typical example of multi-component NH perturbations is $H_1 = \sum_{i=x,y,z} \eta_i \sigma_i $ with momentum-independent parameters $\eta_i$. The corresponding rotation matrix, which aligns the oriented $\sigma'_z$-axis with the direction $(\theta, \varphi)$, is explicitly constructed as 
\begin{equation}
R(\theta,\varphi)= \eta_{xyz}^{-1}\eta_{xy}^{-1}\begin{pmatrix} 
	\eta_x\eta_z & \eta_y\eta_z & -\eta_{xy}^2 \\
	-\eta_y\eta_{xyz} & \eta_x\eta_{xyz} & 0 \\
	\eta_x\eta_{xy} & \eta_y\eta_{xy} & \eta_z\eta_{xy}
	\end{pmatrix},
\end{equation}
where $\eta_{xy} = \sqrt{\eta_x^2+\eta_y^2}$ and $\eta_{xyz} = \sqrt{\eta_x^2+\eta_y^2+\eta_z^2}$. We revisit the NH QAH model with $H_1 = \eta\sum_{i=x,y,z} \sigma_i$.  Figure~\ref{fig2}(b) presents the rotated TASTs $\langle \sigma'_z (\bm{k}) \rangle_T$ with parameters $(m_z,t_{so},\eta)= (3,0.5,0.3)t_0$. The BIS, identified by $\langle\sigma'_z(\bm{k})\rangle_T=0$,  manifests as a ring-shaped structure indicated by the black dashed curve. In Fig.~\ref{fig2}(c), the dynamical field $\bm{g}(\phi)$ along the BIS contour (purple arrows), constructed from the normalized TASTs $\overline{\langle\sigma'_{x,y}(\phi)\rangle}_T$ as functions of the azimuthal angle $\phi$, yields a nonzero winding number $W_1=1$.

\section{Generalization to higher dimensions}\label{HDNH}

Our dynamical measurement protocol can be extended to higher even dimensions. 
We consider $d$D ($d>2$) real line-gapped systems described by the Hamiltonian
\begin{align}\label{HdD}
	\begin{split}
		H({\bm k})&= H_0({\bm k})+\ui H_1({\bm k}),\\
		H_0({\bm k}) &={\sum}_{i=0}^d h_i({\bm k})\gamma_i, \quad H_1(\bm{k}) = h'_0(\bm{k})\gamma_0,
	\end{split}
\end{align}
where $\gamma_i$ are $2^{d/2}\times2^{d/2}$ matrices for an even $d$, the minimal dimension required to open a line gap, satisfying the Clifford algebra $\{\gamma_i,\gamma_j\}=2 \delta_{ij}\id$ ($i,j=0,1,\dots,d$).
The complex eigenenergies are $\pm \varepsilon (\bm{k}) $, with $\varepsilon(\bm{k}) =[\sum_{i=1}^{d} h_i^2+(h_0+\ui{h'_0})^2]^{1/2}$.
The Hamiltonian \eqref{HdD} can generally describe NH phases across multiple symmetry classes that are classified by integer topological invariants.
As demonstrated in Sec.~\ref{II.B}, the scheme can also be applied to general scenarios where multiple NH terms
maintain a fixed orientation in the spinor space spanned by the $\gamma$ matrices. 
Through unitary transformation of the $\gamma$-matrix basis, such systems can always be reduced to the canonical representation given in Eq.~(\ref{HdD}).

Similar to 2D systems, we introduce an auxiliary matrix $Q(\bm{k})=\sum_{i=0}^d h^Q_i(\bm{k})\gamma_i$  as defined in Eq.~\eqref{Q_def}.
To ensure the emergence of chiral symmetry,
the BISs of $Q(\bm{k})$ should be defined along the $\gamma_0$-axis, which correspond to the momentum subspaces where $h^Q_0(\bm{k})=0$ (equivalently, $h_0(\bm{k})=0$; see Appendix~\ref{LocateBISs}).
We note that on the BISs, $h^Q_i(\bm{k}) = h_i(\bm{k})/\varepsilon(\bm{k})$ for $i=1,2,\dots,d$,
where $\varepsilon(\bm{k})=[\sum_{i=1}^d h_i(\bm{k}) ^2-{h'_0}^{2}(\bm{k})]^{1/2}>0$ is strictly real.
This implies $h^Q_i(\bm{k})\approx h_i(\bm{k})$ on BISs. 
For subsequent topological characterization using BISs, we will consistently employ $h_i(\bm{k})$ when no ambiguity arises.
Thus, the topology of the $Q$ matrix is characterized by a $(d-1)$D winding number defined on all BISs:
\begin{align}\label{Wd}
	W_{d-1}=\sum_j\frac{\Gamma(d/2)}{2\pi^{d/2}(d-1)!}\int_{{\rm BIS}_j}{{\hat{\bm{h}}}_{\rm so}}({\rm {d}}{{\hat{\bm{h}}}_{\rm so}})^{d-1},
\end{align}
where $\bm{\hat{h}}_{\rm so}(\bm{k})=(\hat{h}_1(\bm{k}),\hat{h}_2(\bm{k}),...,\hat{h}_d(\bm{k}))$ is the normalized spin-orbit coupling field, with $\hat{h}_i = {h_i}/\sqrt{\sum_{m=1}^d  h_m^2}$,  ${\Gamma}(x)$ is the Gamma function, ${{\hat{\bm{h}}_{\rm so}}}(\ud{{\hat{\bm{h}}_{\rm so}}})^{d-1}\equiv\epsilon^{i_{1}i_{2}\cdots i_{d}}\hat{h}_{i_{1}}\ud \hat{h}_{i_{2}}\wedge\cdots\wedge\ud \hat{h}_{i_{d}}$
with $\epsilon^{i_{1}i_{2}\cdots i_{d}}$ being the fully antisymmetric tensor and $i_{1,2,\dots,d+1}\in\{1,2,\dots,d\}$, and `$\ud$' denotes the exterior derivative.

To characterize the winding number $W_{d-1}$, we further introduce second-order BISs (2BISs) defined by the simultaneous vanishing of two Hamiltonian components.
Without loss of generality, we define 2BISs as
\begin{align}~\label{2BIS_def}
	{\rm 2 BIS}\equiv\{{\bm k}| h_0(\bm{k}) =0\wedge h_{1}({\bm k})=0\}.
\end{align}
Here $h_{1}$ can be replaced by any other component $h_{j\neq0}$ (see Appendix~\ref{measure}).
Through the higher-order bulk-surface duality~\cite{Yu2021},
the winding number $W_{d-1}$ can reduce to a $(d-2)$D Chern number $C_{d-2}$ defined on the 2BISs.
Crucially, the emergent chiral symmetry of the NH Hamiltonian $H({\bm k})$  on BISs 
enables direct extraction of $C_{d-2}$ via quench dynamics protocols as demonstrated in Ref.~\cite{lin2025}.

According to the definition in Eq.~\eqref{2BIS_def}, we choose the initial state to be polarized along the $\gamma_1$ axis, i.e., $\rho_0 = (\id - \gamma_1)/2^{d/2}$.
The location of 2BISs can be dynamically determined by identifying momentum subspaces where (Appendix~\ref{measure})
\begin{equation}
 \langle\gamma_{j}(\bm{k})\rangle_T= 0\,\,\,\textrm{for all $j=2,3,\cdots,d$}.
\end{equation}
We further introduce a dynamical field $\boldsymbol{\mathcal{G}}(\bm{k})=(\mathcal{G}_2,\mathcal{G}_3,\dots,\mathcal{G}_d)$ defined on the 2BISs through
\begin{equation}\label{Gj_def}
	{\mathcal{G}_j(\bm{k})} \equiv -\frac{1}{\mathcal{N}_g}\partial_{k_\perp} {\langle\gamma_{j}(\bm{k})\rangle}_T,\,\,\, j=2,\cdots,d,
\end{equation}
with $\mathcal{N}_g$ being the normalization factor. Here $k_\perp$ is the momentum perpendicular to the contour of 2BISs, pointing from ${h}_1(\bm{k})<0$ to  ${h}_1(\bm{k})>0$ regions on the BISs. 
One can verify that ${\mathcal{G}_j(\bm{k})}|_{{\bm k}\in \mathrm{2BIS}} = \hat{h}_j(\bm{k})$, where $\hat{h}_{j}({\bm k})=h_j(\bm{k})/\sqrt{\sum_{m=2}^d h^2_m(\bm{k})}$ (Appendix~\ref{measure}), which leads to the characterization of the $(d-2)$D Chern number:
\begin{equation}\label{C_dyn}
	\begin{aligned}
%		C_{d-2} &= \sum_j C_{d-2}^{j}, \\
%		C_{d-2}^{j} &= \frac{{\Gamma}\left[(d-1)/2\right]}{2\pi^{(d-1)/2}}\frac{1}{(d-2)!}\int_{{\rm 2BIS}_j}{\boldsymbol{\mathcal{G}}}(\ud{\boldsymbol{\mathcal{G}}})^{d-2}.
C_{d-2} = \sum_j \frac{{\Gamma}\left[(d-1)/2\right]}{2\pi^{(d-1)/2}(d-2)!}\int_{{\rm 2BIS}_j}{\boldsymbol{\mathcal{G}}}(\ud{\boldsymbol{\mathcal{G}}})^{d-2}.
	\end{aligned}
\end{equation}
Therefore, the topology of the $d$D NH Hamiltonian $H(\bm{k})$ can be detected by the $(d-2)$D Chern number 
of the dynamical field $\boldsymbol{\mathcal{G}} (\bm{k})$ defined on 2BISs, which is directly measurable in experiment.

\begin{figure}
	\includegraphics[width=0.48\textwidth]{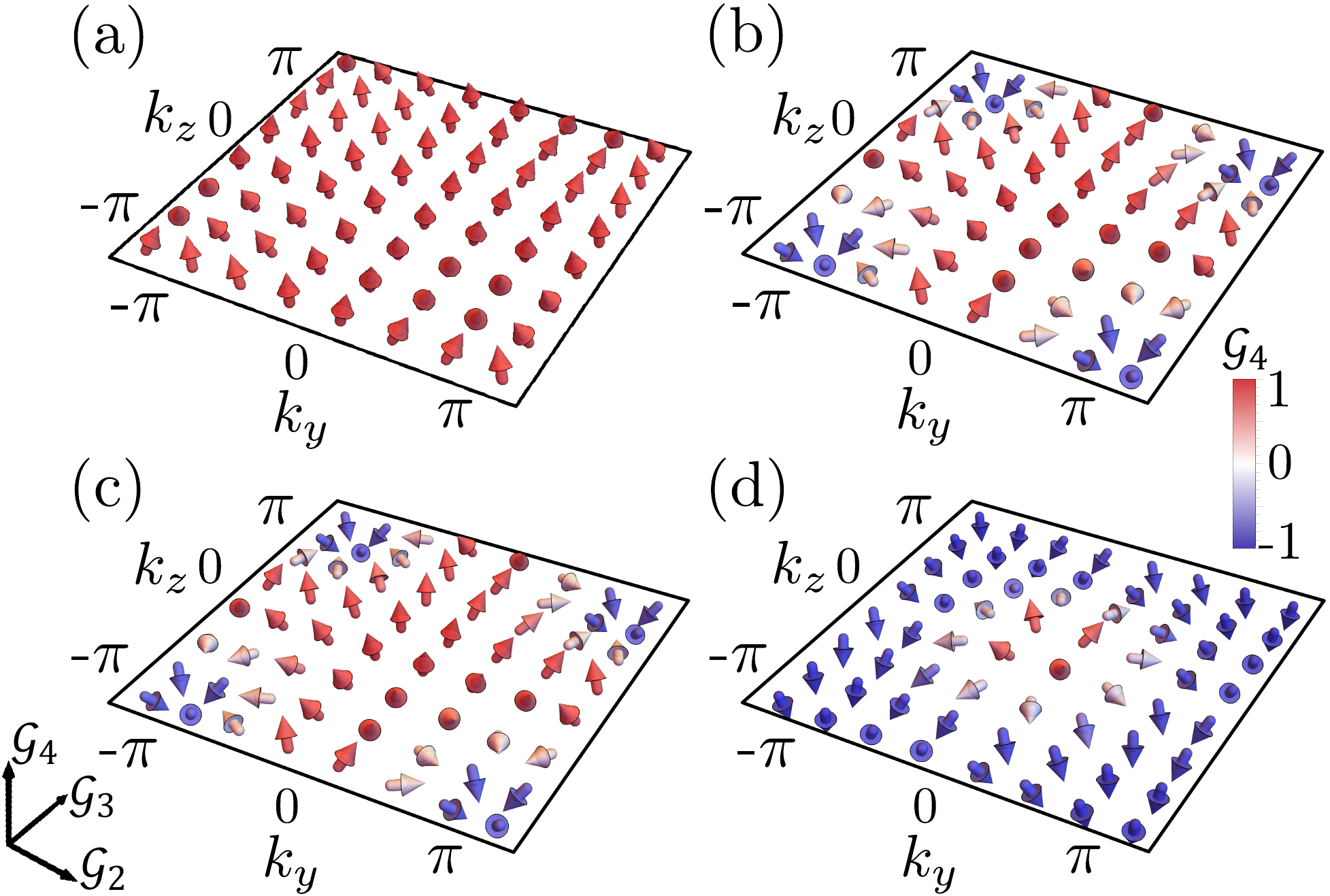}
	\caption{ Measuring the topology of the 4D NH topological insulator. The dynamical field $\bm{\mathcal{G}}(\bm{k})$ is shown as arrows on four 2BISs corresponding to $(k_w, k_x) =(0,0)$ (a), $(0,\pi)$ (b), $(\pi,0)$ (c), and $(\pi,\pi)$ (d). These configurations contribute Chern numbers $(C_2^{(1)},C_2^{(2)},C_2^{(3)},C_2^{(4)})=(0,-1,-1,-1)$~\cite{note2}. Here we take the parameters $(m,\eta)= (1,0.2)$ and the initial state $\rho_0 = (1-\gamma_1)/4$.}\label{fig3}
\end{figure}

Here we take a four-dimensional NH topological insulator as an example, with the Hamiltonian~\cite{Liang2008}:
\begin{align}
\begin{split}
		H_0 =&\sin k_w \gamma_0 + \sin k_x \gamma_1 + \sin k_y \gamma_2 + \sin k_z \gamma_3  \\
		&+\left(m + \cos k_x + \cos k_y + \cos k_z + \cos k_w\right)\gamma_4, \\
		H_1 =&\,\eta \gamma_0,
\end{split}
\end{align}
where the Clifford algebra generators are chosen to be $\bm{\gamma} =\left(\sigma_x \otimes \tau_0, \sigma_y\otimes\tau_0, \sigma_z\otimes\tau_x, \sigma_z\otimes\tau_y, \sigma_z\otimes\tau_z \right)$. Here $\sigma_i$ and $\tau_i$ are both Pauli matrices.
%and the momentum-dependent mass term $m(\bm{k})=$ contains the Dirac mass $m$ and nearest-neighbor hopping amplitude $t_0$. 
The Hermitian part $H_0(\bm{k})$ exhibits topological phases classified by the second Chern number
\begin{equation}
	C_2 = 
	\begin{cases}
		0, &  |m| > 4 \\
		1, & 2 < m < 4 \\
		-3, & 0 < m < 2 \\
		3, & -2 < m < 0\\
		-1, & -4 < m < -2
	\end{cases}.
\end{equation}
For parameters $(m,\eta) = (1, 0.2)$ and the initial state $\rho_0 = (\id - \gamma_1)/4$, we derive the TASTs $\langle \gamma_j(\bm{k}) \rangle_T$ ($j = 2, 3, 4$) and identify momentum planes $(k_w, k_x) = (0, 0)$, $(0, \pi)$, $(\pi, 0)$, $(\pi, \pi)$ as four 2BISs.  
As shown in Figs.~\ref{fig3}(a)-(d), the constructed dynamical field $\boldsymbol{\mathcal{G}}(\bm{k})$ on each 2BIS is represented by vector arrows. 
The vector configurations on the four 2BISs contribute individual Chern numbers $(C_2^{(1)},C_2^{(2)}, C_2^{(3)}, C_2^{(4)}) = (0,-1, -1, -1)$~\cite{note3}, resulting in a total Chern number $C_2 =\sum_{i=1}^4 C_2^{(i)}=-3$.

Before concluding, we reemphasize the dimensional dichotomy in topological characterization between 2D and higher even-dimensional NH systems. 
This distinction originates fundamentally from the dimension-dependent Clifford algebra structure encoded in Eq.~\eqref{HdD}.
Specifically, 2D systems are governed by three Pauli matrices to describe two-band physics, whereas higher even-dimensional systems 
require expanded $\gamma$ matrices of dimension $2^{d/2}\times2^{d/2}$
to stabilize real line gaps. Consequently, the 2D case necessitates carefully constructed initial states polarized along the NH perturbation axis, 
while in higher even dimensions, a generic initial state of the form
$\rho_0 = (\id - \gamma_i)/2^{d/2}$ ($i = 1, 2, \dots, d$) suffices to extract topological invariants from quench dynamics (Appendix~\ref{measure}).

\section{Discussion and conclusions}\label{Conculsion}

In this work, we propose a dynamical characterization scheme for real line-gapped topological phases with integer invariants in even-dimensional NH systems.
Reconstructing full topological information from quench-induced non-unitary dynamics is fundamentally obstructed by inherent gain and loss:
amplifying eigenstates dominate over time, while decaying modes rapidly become unobservable.
Our approach surmounts this difficulty by exploiting momentum-space bulk-surface duality, enabling topological characterization through BISs adapted to NH perturbations,
where emergent chiral symmetry preserves essential topological information.
Specifically, we demonstrate that the topology in 2D NH systems is quantified by the winding number of the dynamical field $\bm{g}(\bm{k})$ along 1D BISs,
while in higher even dimensions ($d > 2$), it is determined by the $(d-2)$D Chern number of the dynamical field $\boldsymbol{\mathcal{G}}(\bm{k})$ on 2BISs.

%In this work, we propose a dynamical scheme to characterize even-dimensional, integer-invariant topological phases in NH systems. The main idea is to exploit the bulk-surface correspondence in momentum space, which can characterize the topology of NH Hamiltonian on the BISs, defined aligned with the NH terms. Specifically, for two-dimensional NH systems, the topology is characterized by the winding number of the dynamical field $\bm{g}(\bm{k})$ along the BISs. In higher even dimensions $d > 2$, the topology is captured by the $(d-2)$D Chern number of the dynamical field $\boldsymbol{\mathcal{G}}(\bm{k})$ on the 2BISs.

The general Hamiltonian in Eq.~\eqref{HdD} universally describes NH systems with real line gaps across multiple symmetry classes,
including complex class A and real classes AI, $\text{AI}^\dagger$, D, $\text{D}^\dagger$, AII, $\text{AII}^\dagger$, C and $\text{C}^\dagger$~\cite{Kawabata2019a}. 
The dynamical characterization scheme can also be extended to even-dimensional systems with global sublattice or chiral symmetry, following the methodology established in Ref.~\cite{lin2025}. 
As demonstrated in Appendix~\ref{NHBHZ} through the 2D NH Bernevig-Hughes-Zhang (BHZ) model with sublattice symmetry, which reduces to the quantum spin Hall effect in the Hermitian limit, a dynamical field constructed from long-time spin textures (LTSTs) can be employed to characterize the topology.
An extension of the scheme to imaginary line-gapped NH systems is presented in Appendix~\ref{imag_line_gap}.

%The general form of Eq.(\ref{HdD}) broadly applies to real line-gapped NH phases across multiple symmetry classes, including complex A class and real AI, $\text{AI}^\dagger$, D, $\text{D}^\dagger$ AII, $\text{AII}^\dagger$, C and $\text{C}^\dagger$ classifications~\cite{Kawabata2019a}. Moreover, the scheme can also be extended to even dimensional systems with global sublattice or chiral symmetry. In Appendix C, we illustrate this by analyzing NH Bernevig-Hughes-Zhang (BHZ) model with sublattice symmetry, which describes the quantum spin Hall effect at the Hermitian limit. In contrast to the scheme presented in the main text, the dynamic field is constructed by LTSTs and the topology is defined on the entire BZ~\cite{lin2025}.

Under open boundary conditions, non-Bloch topological invariants are formulated to characterize the topology through the complex momentum mapping $e^{\ui k}\to\beta$~\cite{Yao2018a}.
The resultant non-Bloch Hamiltonian $H(\beta)$ typically contains NH terms that do not exhibit a fixed orientation in the spinor space, 
obstructing the construction of chiral-symmetric momentum subspaces.
However, in systems with sublattice symmetry, these non-Bloch invariants defined over the generalized Brillouin zone are also directly measurable from quench dynamics~\cite{lin2025}.

The proposed scheme is experimentally feasible with state-of-the-art ultracold atom techniques.
Leveraging existing implementations of 2D Chern insulators in optical Raman lattices~\cite{Zhan2016,Sun2018a,Liang2023}, 
controlled dissipation can be engineered through atomic resonant coupling to short-lived excited states~\cite{Ren2022}.
The quench protocol can be implemented by applying an additional Zeeman field 
and abruptly reducing its strength from a large initial value to zero. 
TASTs for topological characterization are then measured through spin-resolved time-of-flight imaging~\cite{Sun2018b,Yi2019,WangZY2021}.
Crucially, although the TASTs are formally defined in the infinite-time limit,  
experiments~\cite{Sun2018b,Yi2019} demonstrate that {\it short-time} dynamics---spanning only several oscillation periods on BISs---provides sufficient information~\cite{note4}. 
Simultaneously, high spontaneous emission rates ensure excited states act as efficient particle sinks, making quantum jumps negligible within the measurement window. 
Furthermore, boundary effects become insignificant when the atomic cloud remains sufficiently distant from system boundaries throughout its evolution~\cite{Zhao2023}.
Under these conditions, the dynamics is faithfully governed by a NH Bloch Hamiltonian $H(\bm{k})$.
Our results thus establish a practical protocol for experimental detection and characterization of NH topological phases, 
with direct applicability to quantum simulators such as ultracold atomic systems.

\section*{Acknowledgements}

This work was supported by the National Natural Science Foundation of China (Grants No. 12204187),  
the Innovation Program for Quantum Science and Technology (Grant No. 2021ZD0302000), 
and the startup grant of Huazhong University of Science and Technology (Grant No. 3034012114).

\section*{Data Availability}

The data that support the findings of this article are openly available~\cite{Data}.

\appendix

\section{Dynamical identification of band-inversion surfaces}\label{LocateBISs}

In this Appendix, we demonstrate the identification of BISs through quench dynamics.
Starting from the general NH Hamiltonian in Eq.~(\ref{HdD}) and the corresponding auxiliary matrix $Q(\bm{k})=\sum_{i=0}^d h^Q_i(\bm{k})\gamma_i$,
we first prove that the defining condition $h^Q_0(\bm{k}) = 0$ is equivalent to $h_0(\bm{k}) = 0$, both specifying identical momentum subspaces.

Our proof begins with the definition in Eq.~\eqref{Q_def}, which leads to
\begin{equation}\label{h^Q_0}
	h^Q_0(\bm{k}) = \frac{1}{2} \left[ \frac{h_0(\bm{k}) + \ui h'_0(\bm{k})}{\varepsilon(\bm{k})} + \frac{h_0(\bm{k}) - \ui h'_0(\bm{k})}{\varepsilon^*(\bm{k})} \right],
\end{equation}
where $\varepsilon(\bm{k}) =\sqrt{\sum_{i=1}^d h_i(\bm{k}) ^2+[h_0(\bm{k})+\ui{h'_0}(\bm{k})]^2}$.
We rewrite $\varepsilon(\bm{k}) = |\varepsilon(\bm{k})| e^{\ui \theta(\bm{k})}$ and substitute it into Eq.~\eqref{h^Q_0}, yielding
\begin{equation}
	h^Q_0(\bm{k}) = \frac{1}{|\varepsilon(\bm{k})|} \left[ h_0(\bm{k}) \cos \theta(\bm{k}) + h'_0(\bm{k}) \sin \theta(\bm{k}) \right].
\end{equation}
Thus, the defining condition $h^Q_0(\bm{k}) = 0$ is equivalent to
\begin{equation}
	h_0(\bm{k}) = - h'_0(\bm{k}) \tan \theta(\bm{k}).
	\label{eq:h0_relation}
\end{equation}
Since $ \varepsilon^2(\bm{k}) = |\varepsilon(\bm{k})|^2 e^{\ui 2\theta(\bm{k})} = \sum_{i=0}^d h_i^2(\bm{k}) - h_0'^2(\bm{k}) + 2 \ui h_0(\bm{k}) h'_0(\bm{k})$, we have
\begin{equation}
	\tan 2\theta(\bm{k}) = \frac{2 h_0(\bm{k}) h'_0(\bm{k})}{\sum_{i=0}^d h_i^2(\bm{k}) - h_0'^2(\bm{k})}.
	\label{eq:tan2theta}
\end{equation}
Equation \eqref{eq:h0_relation} requires
\begin{equation}\label{tan2theta}
	\begin{aligned}
		\tan 2\theta(\bm{k}) = \frac{2 \tan \theta(\bm{k})}{1 - \tan^2 \theta(\bm{k})}=\frac{2 h_0(\bm{k}) h'_0(\bm{k})}{ h_0^2(\bm{k}) - h_0'^2(\bm{k})}.
	\end{aligned}
\end{equation}
Comparing Eqs.~\eqref{eq:tan2theta}  and \eqref{tan2theta} yields the equivalent condition for defining the BISs:
\begin{equation}
	h_0(\bm{k}) = 0,
\end{equation}
or
\begin{equation}
	\tan \theta(\bm{k}) = 0.
\end{equation}
The second formulation implies that eigenenergies become purely real on the BISs.

We then demonstrate that the BISs can be dynamically identified by the vanishing of TASTs regardless of the initial state. For an initial state $\rho_0$, we examine the TAST
\begin{equation}
	\langle\gamma_0(\bm{k})\rangle_T = \lim_{T\to\infty} \frac{1}{T} \int_0^T dt\, \frac{1}{\mathcal{N}_{\bm{k}}} \mathrm{Tr}\left[\rho_0(\bm{k}, t) \gamma_0\right],
\end{equation}
with the normalization factor ${\cal N}_{\bm{k}}\equiv{\rm Tr}\left[\rho_0(\bm{k},t)\right]$.
On the BISs, one can derive that
\begin{align}~\label{gamma_0}
		{\rm Tr}\left[\rho_0(\bm{k},t)\gamma_0\right] =& {\rm Tr}\left(\rho_0 e^{\ui H^{\dagger}t}\gamma_0e^{-\ui H t}\right) \nonumber\\
		=& {\rm Tr}\left(\rho_0\gamma_0 e^{-\ui H t} e^{-\ui H t}\right) \nonumber\\
		=& {\rm {Tr}} \left[ \rho_0 \gamma_0\left(\cos 2\varepsilon t -\ui \frac{\sin 2\varepsilon t}{\varepsilon}\sum_{i=1}^d h_i\gamma_i\right.\right.\nonumber\\
		&\left.\left.+ \frac{h'_0\gamma_0}{\varepsilon} \sin 2\varepsilon t\right)\right].
\end{align}
%and 
%\begin{align}
%		{\cal N}_{\bm{k}} =& {\rm Tr}\left\{ \rho_0 \left[1 + \frac{h'_0\gamma_0}{\varepsilon} \sin 2\varepsilon t+ \frac{1-\cos2\varepsilon t}{\varepsilon^2}\right.\right. \nonumber \\
%		&\left.\left.\times\left({h'_0}^{2}-\ui h'_0\gamma_0\sum_{i=1}^d h_i\gamma_i\right)\right]\right\},
%\end{align}
From Eq.~\eqref{gamma_0}, we see that regardless of $\rho_0$, the spin polarization oscillates persistently around zero, resulting in $\langle\gamma_0(\bm{k})\rangle_T = 0$ on the BISs. In regions away from BISs, the energy spectra acquire complex eigenvalues, with only the right eigenvectors corresponding to positive imaginary energy components surviving the long-time evolution. Consequently, the spin polarization decays rapidly, causing the TAST $\langle\gamma_0(\bm{k})\rangle_T$ asymptotically approaching non-vanishing expectation values determined by the dominant right eigenstates. This stark contrast in dynamical responses---persistent oscillations on BISs versus rapid relaxation elsewhere---provides a robust experimental signature for identifying BISs.

\section{Details on the construction of dynamical fields}\label{measure}

In this subsection, we provide technical details on the formulation of dynamical fields for determining NH topological invariants.
On the BISs, the NH Hamiltonian takes the form
\begin{align}~\label{Hd_1D}
	\begin{split}
		H({\bm k})& = {\sum}_{i=1}^d h_i({\bm k})\gamma_i+\ui h'_0(\bm{k})\gamma_0,
	\end{split}
\end{align}
which has the chiral symmetry $\Gamma=\gamma_0$, defined by $\Gamma H^\dag({\bm k})\Gamma^{-1}= -H({\bm k})$~\cite{Kawabata2019a}.
The TASTs $\langle\gamma_i(\bm{k})\rangle_T$ ($i = {1,2,...,d}$) on the BISs for an initial state $\rho_0$ read as
\begin{align}
	\langle\gamma_i(\bm{k})\rangle_T=\lim_{T\to \infty}\int_0^T\frac{1}{{\cal N}_{\bm{k}}}{\rm Tr}\left[\rho_0(\bm{k},t)\gamma_i\right] dt,
\end{align}
where the integrand is given by
\begin{align}\label{integrand}
		{\rm Tr}\left[\rho_0(\bm{k},t)\gamma_i\right] =&{\rm Tr}\left[\rho_0\gamma_i\left(\cos^2 \varepsilon t - \ui \sin2\varepsilon t \frac{\sum_{m=1}^d h_m\gamma_m}{\varepsilon}\right.\right.\nonumber\\
		& -\sin^2\varepsilon t \frac{\varepsilon^2 - 2\ui {h'}_{0} \gamma_0 \sum_{m=1}^d h_m\gamma_m + {h'}_{0}^2}{\varepsilon^2}\nonumber\\
		&\left.\left.+ \ui \sin2\varepsilon t \frac{h_i \gamma_i}{\varepsilon} + 2\sin^2\varepsilon t \frac{h_i \gamma_i H }{\varepsilon^2} \right)\right].
\end{align}
It should be noted that Eq.~\eqref{integrand} assumes this simple form precisely because of the emergent chiral symmetry on the BISs.
%Additionally, we note that on the BISs,
%\begin{align}
%h^Q_i(\bm{k}) = \frac{h_i(\bm{k})}{\varepsilon(\bm{k})}, \quad i=1,2,\dots,d,
%\end{align}
%where $\varepsilon(\bm{k})=\sqrt{\sum_{i=1}^d h_i(\bm{k}) ^2-{h'_0}^{2}(\bm{k})}>0$ is strictly real.
%This implies $h^Q_i(\bm{k})\approx h_i(\bm{k})$ on BISs. 
%For subsequent topological characterization using BISs, we will consistently employ $h_i(\bm{k})$ when no ambiguity arises.
Since $h^Q_i(\bm{k})\approx h_i(\bm{k})$ holds on the BISs, we will consistently use $h_i(\bm{k})$ for subsequent topological characterization based on BISs.

We first consider higher-dimensional systems with $d\geq 4$ and set the initial state to be $\rho_0=(\id-\gamma_s)/2^{d/2}$ ($s\neq0$). 
Accordingly, the 2BISs are defined as
\begin{align}~\label{2BIS_def_geneal}
	{\rm 2 BIS}\equiv\{{\bm k}| h_0(\bm{k}) =0\wedge h_{s}({\bm k})=0\}.
\end{align}
The expression for TASTs on the BISs is simplified to
\begin{equation}~\label{ATSTCS}
	{\langle\gamma_j(\bm{k})\rangle}_T =-\alpha\frac{h_s({\bm k})h_j({\bm k})}{\varepsilon^2(\bm{k})},\quad j\neq0,s.
\end{equation}
where $\alpha\equiv\lim_{T\to \infty}\frac{1}{T}\int_0^T\frac{1}{{\cal N}_{\bm{k}}} (1-\cos 2\varepsilon t)$ is positive. 
From Eq.~\eqref{ATSTCS}, one can easily conclude that the 2BISs defined in Eq.~\eqref{2BIS_def_geneal} can be dynamically identified by the vanishing of TASTs ${\langle\gamma_j(\bm{k})\rangle}_T$ for all $j\neq0,s$.
%We then define the second-order BISs~\cite{Yu2022}:
%\begin{align}~\label{BIS_def}
%	{\rm 2 BIS}\equiv\{{\bm k}| h_0(\bm{k}) \wedge h_i({\bm k})=0\}. 
%\end{align} 
Along 2BISs, the momentum ${\bm k}$ can be decomposed into orthogonal components $(k_{\perp},\bm{k}_{\parallel})$, where $k_{\perp}$ ($\bm{k}_{\parallel}$) corresponds geometrically to the component perpendicular (parallel) to the contour of 2BISs.
To characterize the topology, we introduce a dynamical spin-texture field $\boldsymbol{\mathcal{G}} (\bm{k})=(\mathcal{G}_1,\mathcal{G}_2,\dots,\mathcal{G}_{s-1}, \mathcal{G}_{s+1},\dots,\mathcal{G}_d)$, whose components are given as in Eq.~\eqref{Gj_def}.
We mathematically posit that in the vicinity of 2BISs, the function $h_s(\bm{k})$ exhibits monotonic dependence and is uniquely parameterized by the perpendicular momentum component $k_{\perp}$.
One can derive from Eq.~\eqref{ATSTCS} that~\cite{Zhang2018}
\begin{align}~\label{G=Q}
	{\mathcal{G}_j(\bm{k})}\vert_{{\bm k}\in \mathrm{2BIS}} =& \frac{\alpha}{\mathcal{N}_g}\cdot\frac{1}{2k_{\perp}}\frac{2k_{\perp}\cdot h_{j}(0,\bm{k}_{\parallel})}
{\sum_{m=2}^d h^2_m(0,\bm{k}_{\parallel})}+\mathcal{O}(k_{\perp}) \nonumber\\
  =&\hat{h}_j(\bm{k}).
\end{align}
This leads to the dynamical characterization of the NH topology as formulated in Eq.~\eqref{C_dyn}.

Unlike their higher-dimensional counterparts, 2D NH systems require initial state polarization along the NH perturbation axis (i.e., the emergent chiral symmetry axis): $\rho_0 = (\id+\sigma_z)/{2}$~\cite{lin2025}. We then have
\begin{align}~\label{TAST_1D}
	{\langle\sigma_x\rangle}_T= -\alpha_1 \frac{h_y}{\varepsilon^2}, \quad {\langle\sigma_y\rangle}_T = \alpha_1 \frac{h_x}{\varepsilon^2},
\end{align}
where 
\[
\alpha_1\equiv\lim_{T\to \infty}-\frac{1}{T}\int_{0}^{T}dt \frac{1}{{\cal N}_{k}}\left[2h'_z\sin^2(\varepsilon t)+\varepsilon\sin(2\varepsilon t)\right].
\]
We employ the normalized TASTs
\begin{equation}~\label{Ave_1D}
	\overline{\langle\sigma_i(\bm{k})\rangle}_T=\frac{{\langle\sigma_i(\bm{k})\rangle}_T}{\sqrt{\sum_{j=x,y} \langle\sigma_{j}(\bm{k})\rangle_T^2}} ,\quad i =x,y,
\end{equation}
and define the dynamical field ${\bm g}(\bm{k})=(g_{x}(\bm{k}),g_{y}(\bm{k}))$ as presented in Eq.~\eqref{g_def}.
It can be derived from Eqs.~\eqref{TAST_1D} and \eqref{Ave_1D} that on the BISs, 
\begin{equation}~\label{2Dgi}
	g_{i}(\bm{k})|_{{\bm k}\in \mathrm{BIS}} = \text{sgn}(\alpha_1)\cdot\hat{h}_i(\bm{k}),\quad i=x,y,
\end{equation}
where $\hat{h}_i(\bm{k}) = h_i(\bm{k})/\sqrt{\sum_{j=x,y} h^2_j(\bm{k})}$ and $\text{sgn}(\cdot)$ denotes the sign function.
The winding of ${\bm g}(\bm{k})$ is independent of the sign of $\alpha_1$.

\section{Non-Hermitian Bernevig-Hughes-Zhang model}\label{NHBHZ}

\begin{figure}
	\includegraphics[width=0.48\textwidth]{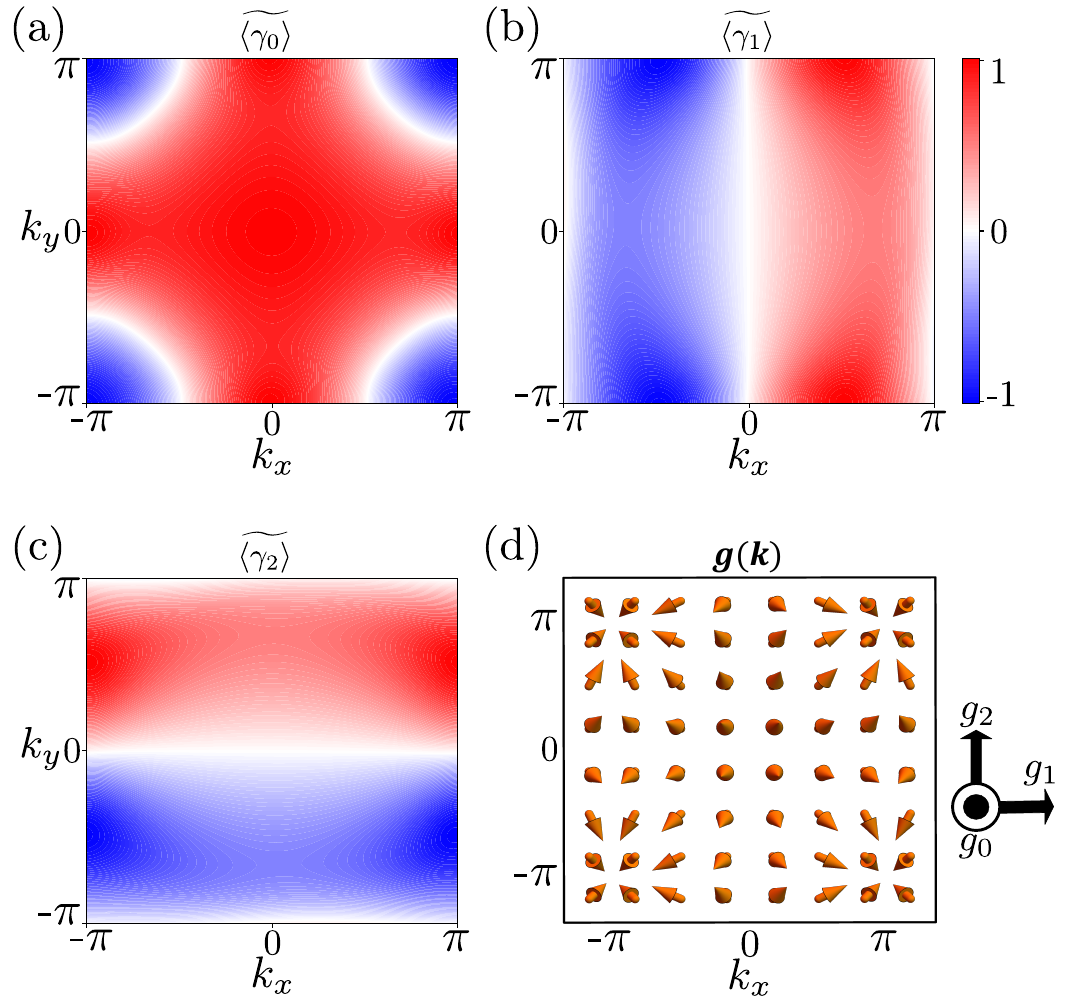}
	\caption{ Measuring the topology of the NH BHZ model.
	(a)-(c) The normalized LTSTs $\widetilde{\langle\gamma_i(\bm{k})\rangle}$. 
	(d) The constructed dynamical field is plotted as arrows throughout the BZ, yielding $C_s = 1$. 
	Here we take the parameters $(m,\eta_0,\eta_1,\eta_2)= (1,0.2,0.2,0.2)t$ and the initial state $\rho_0 = (\id-S)/4$.}\label{fig4}
\end{figure}

In this Appendix, we investigate a NH extension of the BHZ model~\cite{Bernevig2006}, 
distinguished from the 2D model in Sec.~\ref{2DNH} by its global sublattice symmetry: $SH({\bm k})S^{-1}=-H({\bm k})$~\cite{Kawabata2019a}. %The BHZ model describes mercury telluride-cadmium telluride semiconductor quantum wells that host the quantum spin Hall effect, and 
The Hamiltonian is given by 
\begin{align}\label{BHZmodel}
        \begin{split}
        &H({\bm k})=H_0({\bm k})+\ui H_1({\bm k}), \\
	&H_0({\bm k}) = \sum_{i=0}^2 h_i(\bm{k})\gamma_i, \quad H_1(\bm{k}) = \sum_{i=0}^{2} \eta_i\gamma_i,
	\end{split}
\end{align}
where $h_0 = m + t\cos k_x + t\cos k_y$, $h_1 = t\sin k_x$, $h_2 = t\sin k_y$, and the Clifford matrices $(\gamma_0,\gamma_1,\gamma_2 )=( \sigma_0 \otimes \tau_z, \sigma_z \otimes \tau_x, \sigma_0 \otimes \tau_y)$.
%Here $\sigma_{x,y,z}$ ($\tau_{x,y,z}$) denote the Pauli matrices acting on spin (orbital) degrees of freedom. 
%The systems exhibits time-reversal symmetry $\mathcal{T} =  \ui\sigma_y$ and sublattice symmetry $S = \sigma_y \otimes \tau_x$. 
This system has the sublattice symmetry $S = \sigma_y \otimes \tau_x$.
The Hermitian model $H_0$ exhibits topological phases classified by the spin Chern number $C_s\in\mathbb{Z}$~\cite{Sheng2005,Sheng2006},
with nontrivial topology emerging when $0<|m|<2t$,  where $C_s=\text{sgn}(m)$.
These topological phases persist under the non-Hermitian perturbation $H_1$ as long as the real line gap remains open~\cite{Kawabata2020}.

We show that the dynamical measurement scheme proposed in Ref.~\cite{lin2025} can be directly applied to the NH BHZ model.
The $Q$ matrix defined in Eq.~\eqref{Q_def} corresponding to the Hamiltonian \eqref{BHZmodel} is given by
$Q({\bm k})=\sum_{i=0}^2 h^Q_i({\bm k})\gamma_i$,
where 
\begin{align}
h^Q_i(\bm{k}) = \frac{1}{2}\left(\frac{h_i(\bm{k})+\ui\eta_i}{\varepsilon(\bm{k})}+\frac{h_i(\bm{k})-\ui\eta_i}{\varepsilon^*(\bm{k})} \right),
\end{align}
with $\varepsilon(\bm{k}) = \sqrt{\sum_{i=0}^2 (h_i(\bm{k})+\ui\eta_i)^2}$.
To characterize the topology of $Q(\bm{k})$ matrix, we employ the normalized long-time spin textures (LTSTs)
\begin{equation}\label{LTST_def}
	\overline{\langle\gamma_i(\bm{k})\rangle}=\frac{{\langle\gamma_i(\bm{k})\rangle}_\infty}{\sqrt{\sum_{j=0,1,2} \langle\gamma_j(\bm{k})\rangle_\infty^2}} ,  \quad  i=0,1,2,
\end{equation}
where $\langle\gamma_i(\bm{k})\rangle_\infty=\lim_{t\to \infty}{\cal N}_{\bm{k}^{-1}}{\rm Tr}(e^{-\ui Ht}\rho_0 e^{\ui H^{\dagger}t}\gamma_i)$ with the initial state $\rho_0=(\id - S)/4$, and construct the dynamics field ${\bm g}({\bm k})=(g_0({\bm k}),g_1({\bm k}),g_2({\bm k}))$ by the deformed LTSTs as~\cite{lin2025}
\begin{align}\label{gi_def}
	g_i(\bm{k})= \widetilde{\langle\gamma_i(\bm{k})\rangle}=
	\left\{ \begin{array}{ll}
		\overline{\langle\gamma_i(\bm{k})\rangle} & \mathrm{Im}[\varepsilon(\bm{k})]>0\\
		-\overline{\langle\gamma_i(\bm{k})\rangle} & \mathrm{Im}[\varepsilon(\bm{k})]<0
	\end{array} \right. .
\end{align}
Notably, the construction of ${\bm g}({\bm k})$ presented above is enabled by the sublattice symmetry $S$~\cite{lin2025}.
The spin Chern number can be measured through the dynamical field $\bm{g}(\bm{k})$ via
\begin{equation}
	C_{s} =\frac{1}{4\pi}\int_{{\rm BZ}}{\boldsymbol{{g}}}(\ud{\boldsymbol{{g}}})^{2},
\end{equation}
where the integral is over the first Brillouin zone (BZ), ${\boldsymbol{{g}}}(\ud{\boldsymbol{{g}}})^{2}\equiv\epsilon^{i_{0}i_{1}i_{2}}\bm{g}_{i_{0}}\ud \bm{g}_{i_{1}}\wedge\ud \bm{g}_{i_{2}}$ with $\epsilon^{i_{0}i_{1}i_{2}}$ being the fully antisymmetric tensor and $i_{0,1,2}\in\{0,1,2\}$, and `$\ud$' denotes the exterior derivative. 

As an example, we take the parameters $m=t$ and $\eta_i=0.2t$ ($i=0,1,2$).
The deformed LTSTs $\widetilde{\langle\gamma_i(\bm{k})\rangle}$ ($i=0,1,2$) computed via Eq.~\eqref{gi_def} are shown in Figs.~\ref{fig4}(a)-(c).
The constructed dynamical field $\bm{g}(\bm{k})$, displayed in Fig.~\ref{fig4}(d) as arrows, exhibits nontrivial winding throughout the BZ, yielding the spin Chern number $C_s=1$.
Besides, numerical results also confirm that bulk-surface duality persists in this NH BHZ model. 
Figure 4(a) reveals a ring-shaped structure (white contour) surrounding the $M$ point $(\pi, \pi)$, where $\widetilde{\langle\gamma_0(\bm{k})\rangle}=0$ defines the BIS.
Along this BIS contour, the field ${\bm g}({\bm k})$ also exhibits a nontrivial winding number +1, directly corresponding to the bulk topological invariant $C_s$.

\section{Measuring imaginary line-gapped topology}~\label{imag_line_gap}

\begin{figure}
	\includegraphics[width=0.48\textwidth]{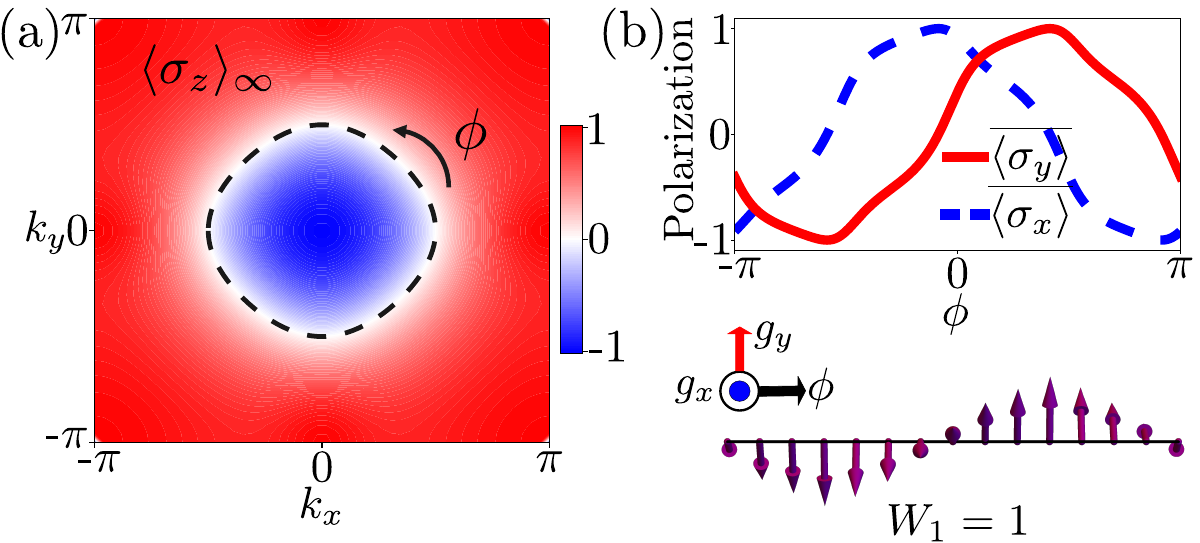}
	\caption{Characterizing the 2D imaginary line-gapped topological phase. (a) A ring-shaped BIS (black dashed curve) is identified via $\langle \sigma_z(\bm{k}) \rangle_\infty = 0$. (b) 
		Upper panel: The normalized LTSTs $\overline{\langle \sigma_{x,y}\rangle}$ along the BIS as functions of the azimuthal angle $\phi$.
		Lower panel: The constructed dynamical field $\bm{g}(\phi)$ (purple arrows) exhibits a nontrivial winding $W_1 = 1$. Here we take the parameters $(m,\eta)=(1,0.4)$ and the initial state $\rho_0 = (\id-\sigma_z)/2$.}\label{fig5}
\end{figure}

In this Appendix, we demonstrate that our dynamical characterization method can also be extended to imaginary line-gapped topological phases,
described by the generic $d$D Hamiltonian
\begin{align}\label{HdD_imag}
	\begin{split}
		H({\bm k})&= H_0({\bm k})+\ui H_1({\bm k}),\\
		H_0({\bm k}) &= h_0({\bm k})\gamma_0, \quad H_1(\bm{k}) = {\sum}_{i=0}^d h'_i(\bm{k})\gamma_i.
	\end{split}
\end{align} 
The complex eigenenergies are $E_{\pm}({\bm k})=\pm \ui \varepsilon'({\bm k})$, with $\varepsilon'\equiv\sqrt{\sum_{i=1}^d h'^2_i+(h'_0-\ui h_0)^2}$.
The associated $Q$ matrix $Q({\bm k})=\sum_{i=0}^d h^Q_i({\bm k})\gamma_i$, as defined in Eq.~\eqref{Q_def}, reads
\begin{align}
	\begin{split}
	h^Q_0(\bm{k}) &= \frac{1}{2}\left(\frac{h'_0(\bm{k})-\ui h_0(\bm{k})}{\varepsilon'(\bm{k})}+\frac{h'_0(\bm{k})+\ui h_0(\bm{k})}{\varepsilon'^*(\bm{k})} \right), \\
	h^Q_i(\bm{k}) &= \frac{1}{2}\left(\frac{h'_i(\bm{k})}{\varepsilon'(\bm{k})}+\frac{h'_i(\bm{k})}{\varepsilon'^*(\bm{k})} \right), \quad i=1,2,...,d .
	\end{split}
\end{align}
	
To characterize the topology of $Q(\bm{k})$, we investigate the LTSTs $\langle\gamma_i(\bm{k})\rangle_\infty$ following a quench from the initial state $\rho_0 = (\id-\gamma_0)/2^{d/2}$.
Further calculations yield
\begin{equation}
\langle\gamma_0(\bm{k})\rangle_\infty =\frac{h^Q_0(\bm{k})}{ \mathcal{N}_{\bm{k}}}\left(2-\frac{4h'_0(\bm{k}) }{\varepsilon'(\bm{k})+\varepsilon'^*(\bm{k})}\right),
\end{equation}
leading to the dynamical characterization of the BISs of $Q(\bm{k})$ (defined by $h^Q_0(\bm{k}) =0$, or equivalently, $h'_0(\bm{k}) =0$):
\begin{equation}
	\langle\gamma_0(\bm{k})\rangle_\infty=0. \label{BISs_imag}
\end{equation}
One can verify that on the BISs, $h^Q_i(\bm{k})\approx h'_i(\bm{k})$ and the LTSTs $\langle\gamma_{i>0}(\bm{k})\rangle_\infty$ take a simple form
\begin{align}
	\langle\gamma_i(\bm{k})\rangle_\infty = \frac{1}{\mathcal{N}_{\bm{k}}}\frac{h'_i(\bm{k})}{\varepsilon'(\bm{k})},\quad i=1,2,...,d,
\end{align}
for higher dimensions with $d\geq 4$. For 2D systems with $H_0(\bm{k})= h_z (\bm{k}) \sigma_z$ and $H_1(\bm{k})=\sum_{i=x,y,z} h'_i (\bm{k}) \sigma_i$,
the LTSTs read 
\begin{align}
	\begin{split}
	& \langle\sigma_x(\bm{k})\rangle_\infty =  \frac{1}{\mathcal{N}_{\bm{k}}} \frac{2}{\varepsilon'(\bm{k})}\left(h'_x(\bm{k})-\frac{h_z(\bm{k})}{\varepsilon'(\bm{k})}h'_y(\bm{k})\right), \\
	& \langle\sigma_y(\bm{k})\rangle_\infty =  \frac{1}{\mathcal{N}_{\bm{k}}} \frac{2}{\varepsilon'(\bm{k})}\left(h'_y(\bm{k})+\frac{h_z(\bm{k})}{\varepsilon'(\bm{k})}h'_x(\bm{k})\right)
	\end{split}
\end{align}
on the BISs. 
Accordingly, we construct the dynamical field $\bm{g}(\bm{k})=(g_1,g_2,...,g_d)$ by the normalized LTSTs $\overline{\langle\gamma_i(\bm{k})\rangle}={\langle\gamma_i(\bm{k})\rangle}_\infty/\sqrt{\sum_{j=1}^d \langle\gamma_j(\bm{k})\rangle_\infty^2}$ as
\begin{align}
	g_i(\bm{k}) = \overline{\langle\gamma_i(\bm{k})\rangle},  \quad  i=1,2,...,d, \label{g_i_imag}
\end{align}
which satisfies $g_i(\bm{k})\vert_{{\bm k}\in\mathrm{BIS}}\approx\hat{h}'_i(\bm{k})$ with $\hat{h}'_i = {h'_i}/\sqrt{\sum_{m=1}^d  h'^2_m}$, 
Thus, the topology of $Q(\bm{k})$ can be dynamically measured through the winding of $\bm{g}(\bm{k})$ on all BISs:
\begin{align}\label{Wd_imag_measure}
	W_{d-1}=\sum_j\frac{\Gamma(d/2)}{2\pi^{d/2}(d-1)!}\int_{{\rm BIS}_j}{{{\bm{g}}}}({\rm {d}}{{{\bm{g}}}})^{d-1}.
\end{align}
Equations (\ref{BISs_imag}) and (\ref{Wd_imag_measure}) provide a direct measurement of the imaginary line-gapped topology of $H(\bm{k})$.
Note that the dynamical characterization is enabled not by chiral symmetry, 
but by pseudo-Hermiticity, $\gamma_0H^\dagger({\bf k}) \gamma_0^{-1}=H({\bf k})$, when restricted to the BISs.

As a concrete example, we consider the 2D NH model
\begin{align}
	 \begin{split}
	&H_0 = \eta \sigma_z, \\
	&H_1 = \sum_{i=x,y}\sin k_i \sigma_i  + (m-\cos k_x -\cos k_y)\sigma_z,
	\end{split}
\end{align}
where $H_1$ supports topologically nontrivial phases within the parameter regime $0<|m|<2$, characterized by the Chern number $C={\rm sgn}(m)$. Figure~\ref{fig5}(a) presents the LTSTs $\langle \sigma_z(\bm{k}) \rangle_\infty$ with parameters $m=1$ and $\eta=0.4$. The BIS, identified by $\langle \sigma_z(\bm{k}) \rangle_\infty=0$, manifests as a ring-shaped structure indicated by the black dashed curve. In Fig.~\ref{fig5}(b), the dynamical field $\bm{g}(\phi)$ (purple arrows), constructed from the normalized LTSTs $\overline{\langle \sigma_{x,y}\rangle}$ along the BIS contour as functions of the azimuthal angle $\phi$, yields a nonzero winding number $W_1=1$, confirming the predicted Chern number under small $\eta$.


\begin{thebibliography}{99}
\bibitem{note1} Throughout this work, we assume the initial state maintains a uniform quasimomentum distribution and is explicitly represented as a spin state for simplicity.
\bibitem{note2} Throughout this work, the symbol ``$\approx$" denotes topological equivalence rather than exact equality. Here, the component $g_i(k)$ is proportional to $\hat{h}^Q_i(k)$ on the BISs, differing primarily by an overall prefactor $\mathrm{sgn}(\alpha_1)$ [see Eq.~(\ref{2Dgi})], which does not impact the winding number.
\bibitem{note3} If we simply look at the dynamical field configurations, we find that the Chern numbers corresponding to Figs.~\ref{fig3}(b) and (c) are both $+1$. However, it is important to note that when integrating the dynamical field over all BISs (similarly for 2BISs), we need to assign a positive orientation to each BIS. For instance, for a 1D BIS, the counterclockwise direction along the region where $h_0<0$ can be defined as the positive orientation [e.g. Fig.~\ref{fig1}(a)]. Consequently, if the integration along each direction is performed directly over $(-\pi, \pi]$, the integration paths for Figs.~\ref{fig3}(b) and (c) would run opposite to the positive orientation. Therefore, when calculating their contributions to the total Chern number, we need to multiply the direct integration result by an additional factor of $-1$.
\end{thebibliography}
\end{document}